\documentclass[
  aps,
  prd,                 
  twocolumn,           
  superscriptaddress,  
  nofootinbib,         
  amsmath,amssymb,
  longbibliography,     
]{revtex4-2}

\usepackage{graphicx}
\usepackage{dcolumn}
\usepackage{bm}
\usepackage{amsthm}
\usepackage{mathrsfs}
\usepackage{xcolor}
\usepackage{tikz}
\usetikzlibrary{positioning, arrows.meta, calc}

\usepackage[utf8]{inputenc}
\usepackage[T1]{fontenc}
\usepackage{etoolbox}
\usepackage{hyperref}
\hypersetup{
  colorlinks=true,
  linkcolor=blue,
  citecolor=blue,
  urlcolor=blue
}

\DeclareMathAlphabet{\mathcal}{OMS}{cmsy}{m}{n}

\theoremstyle{plain} 

\theoremstyle{definition} 

\begin{document}

\preprint{APS/PRD-2025-DRAFT}

\title{Fisher Information Velocity: A New Geometric Channel for Precision Glitch Identification in Gravitational-Wave Detectors}

\author{James Kennington}
 \email{jwkennington@psu.edu}
 \affiliation{Department of Physics, The Pennsylvania State University, University Park, Pennsylvania 16802, USA}
 \affiliation{Institute for Gravitation and the Cosmos, The Pennsylvania State University, University Park, Pennsylvania 16802, USA}

\author{Zach Yarbrough}
 \affiliation{Department of Physics \& Astronomy, Louisiana State University, Baton Rouge, Louisiana 70803, USA}

\date{\today}

\begin{abstract}
Gravitational-wave detectors operate in inherently non-stationary environments, requiring robust detector characterization (DetChar) to distinguish instrumental transients from astrophysical signals. 
Traditional DetChar frameworks typically rely on morphological classifiers or energy-based projections, such as band-limited root-mean-square (BLRMS) metrics, which can conflate global amplitude scaling with physical reconfigurations of the spectrum. 
In this work, we introduce Fisher information velocity, a novel geometric channel that models the detector's power spectral density (PSD) as a point on a Riemannian manifold. 
By tracking the kinematic drift of the noise floor and utilizing exterior algebra to calculate tangent divergence ($\sin \theta$), we mathematically decouple simple energy surges from spectral warps, or differential redistributions of power across frequency bands. 
Applying this framework via the \texttt{sgn-drift} streaming pipeline to ${\sim}40$ hours of high-cadence Advanced LIGO O4a data, we evaluate $N=282{,}080$ independent manifold velocity samples. 
High-resolution phase space mapping reveals a bimodal taxonomy of severe instrumental non-stationarity, classifying events into structural pivots ($87.2\%$) and isotropic surges ($12.8\%$).
Among co-detected events, the geometric channel achieves higher significance than standard BLRMS monitors in $74\%$ of cases with a median sensitivity ratio of $\Gamma = 1.65$.
The two channels detect largely non-overlapping populations, increasing the total anomaly catalog by $87\%$ over BLRMS alone.
Systematic validation on 10 confirmed GWTC-4.0 events and ${\sim}5{,}000$ simulated injections demonstrates robust insensitivity to astrophysical signals, establishing this geometric channel as a sensitive, complementary, and veto-safe diagnostic for current and next-generation detector networks.
\end{abstract}

\maketitle

\section{Introduction}
\label{sec:introduction}

The detection of gravitational waves relies fundamentally on the principle of matched filtering \cite{wainstein_extraction_1970, finn1992detection, allen_findchirp_2012}, which assumes a stationary Gaussian noise background to construct optimal detection statistics \cite{cutler_gravitational_1994, thrane_introduction_2019}.
In the operational regime of terrestrial interferometers, such as Advanced LIGO \cite{TheLIGOScientificCollaboration2015Advanced}, Virgo \cite{Acernese2015Advanced}, and KAGRA \cite{Akutsu2021Overview}, this stationary assumption is routinely violated \cite{Kennington2026_GTSP0, KenningtonBlack2026_GTSP1, KenningtonBlack2026_GTSP2}.
The instrumental noise floor evolves continuously over time due to a complex interplay of physical factors \cite{martynovSensitivityAdvancedLIGO2016, buikemaSensitivityPerformanceAdvanced2020, capote_advanced_ligo_o4_2025}.
Environmental coupling, microseismic variations \cite{matichard2015seismic}, thermal drifts in the suspension systems \cite{saulson1990thermal}, and scattered light transients \cite{soniReducingScatteredLight2021} cause the PSD to fluctuate dynamically.

This non-stationarity introduces severe systemic impacts on both online and offline search pipelines \cite{sachdev_gstlal_2019, nitz_pycbc_2018, ewingPerformanceLowlatencyGstLAL2024}.
First, static whitening models rapidly lose their optimality as the true noise floor drifts away from the reference state, directly reducing the recovered signal-to-noise ratio of astrophysical sources \cite{biscoveanu_quantifying_2020, driggersImprovingAstrophysicalParameter2019, Kennington2026_GTSP0}.
Second, rapid localized shifts in the spectrum manifest as transient noise artifacts, commonly known as glitches.
These non-Gaussian excursions mimic the morphology of compact binary coalescences and unmodeled bursts, artificially inflating the false alarm rate and degrading the statistical confidence of true detections \cite{abbottCharacterizationTransientNoise2016, littenberg_separating_2010}.

To mitigate these effects, detector characterization frameworks actively track the evolution of the detector's spectral density to generate data quality flags and vetoes \cite{abbottGuideLIGODetector2020, davisLIGODetectorCharacterization2021, soni_ligo_detchar_o4a_2025, essickiDQStatistical2021, smithHierarchicalVeto2011}.
However, the efficacy of existing methods is fundamentally limited by how they project and interpret this high-dimensional spectral data \cite{davisLIGODetectorCharacterization2021}.
Morphological projections, such as those utilized by GravitySpy \cite{zevinGravitySpyIntegrating2017} and pipeline-informed noise characterization tools \cite{yarbrough_pinch_2025}, flatten the time-frequency data into two-dimensional spectrograms \cite{davisLIGODetectorCharacterization2021}.
While these image-based methods are highly effective for visual categorization and pattern recognition, they often over-split or under-split categories based on transient visual shape rather than the underlying physical origin of the instrumental disturbance \cite{abbottCharacterizationTransientNoise2016}.

Conversely, energy-based projections, such as band-limited root-mean-square (BLRMS) monitors \cite{davis_improving_2019} and excess-power trigger generators such as Omicron \cite{robinetOmicronTool2020}, track absolute acoustic power across discrete frequency bins or time-frequency tiles.
Because each bin or tile measures power magnitude independently, these methods can conflate global amplitude scaling with structural reconfigurations of the spectrum, as both phenomena produce identical increases in band-limited power \cite{davisLIGODetectorCharacterization2021, abbottCharacterizationTransientNoise2016}.

To resolve this physical degeneracy, the core analytical objective of this work is to mathematically distinguish broad magnitude scaling from spectral warps, events in which power is differentially redistributed across frequency bands, altering the ratios between them.
We propose a novel geometric channel that tracks the continuous relative evolution of the detector state by modeling the space of all admissible PSDs as a Riemannian manifold \cite{Kennington2026_GTSP0, KenningtonBlack2026_GTSP1, KenningtonBlack2026_GTSP2, lee1997riemannian}.
As the physical environment of the interferometer fluctuates, its spectral state traces a continuous trajectory across this geometric space \cite{martynovSensitivityAdvancedLIGO2016, buikemaSensitivityPerformanceAdvanced2020}.

By equipping this statistical manifold with the Fisher information metric \cite{raoInformationAccuracyAttainment1945, amariMethodsInformationGeometry2000}, we assign a formal geometric distance to spectral fluctuations (Eq.~\ref{eq:fisher_metric}) \cite{georgiouDistancesPowerSpectral2006}.
This geometric framework allows us to calculate the Fisher information velocity of the noise floor, mathematically decoupling true physical regime shifts from simple amplitude fluctuations (Eq.~\ref{eq:fisher_velocity}) \cite{Kennington2026_GTSP0, KenningtonBlack2026_GTSP1, KenningtonBlack2026_GTSP2}.
In this work, we demonstrate how continuous kinematic tracking provides the physical depth required to unmask severe instrumental anomalies that are poorly resolved by standard scalar metrics \cite{davis_improving_2019, davisLIGODetectorCharacterization2021}.
Furthermore, we translate this theoretical formalism into an operational streaming architecture, revealing a bimodal taxonomy of instrumental non-stationarity that physically separates simple energy surges, or proportional amplitude scalings that preserve the spectral shape, from pivots, or events with significant spectral warp.

\section{Background}
\label{sec:background}

To quantify the distance between distinct operational states, we model the space of all admissible PSDs as a Riemannian manifold, $\mathcal{M}$ \cite{lee1997riemannian, doCarmo1992riemannian, KenningtonBlack2026_GTSP1}.
As the physical environment of the interferometer fluctuates, its instrumental state traces a continuous kinematic trajectory, $\gamma(t)$, across this high-dimensional geometric manifold \cite{nakahara_geometry_2005, martynovSensitivityAdvancedLIGO2016, buikemaSensitivityPerformanceAdvanced2020}.
We equip this statistical manifold with the Fisher information metric (Eq.~\ref{eq:fisher_metric}) \cite{raoInformationAccuracyAttainment1945, amariMethodsInformationGeometry2000, cover2006elements}.
Because the metric evaluates squared fractional deviations $(\delta S / S)^2$, equal absolute perturbations carry greater geometric weight in quiet spectral regions than in loud ones \cite{georgiouDistancesPowerSpectral2006, jiangGeodesicCurvesManifold2012, barbaresco2013information, helstrom_statistical_1968, cutler_gravitational_1994}.
For a stationary Gaussian process $n(t)$ \cite{papoulis1991probability, creighton2011gravitational, maggiore2007gravitational}, the power spectral density (PSD), $S(f)$, is defined via the expectation value of the Fourier transform \cite{allen_findchirp_2012}:
\begin{equation}
\langle \tilde{n}(f) \tilde{n}^*(f') \rangle = \frac{1}{2} S(f) \delta(f-f').
\label{eq:psd_definition}
\end{equation}
For infinitesimal spectral perturbations $\delta S_1$ and $\delta S_2$, the Fisher metric on the one-sided PSD is expressed as:
\begin{equation}
g_S(\delta S_1, \delta S_2) = \frac{1}{2} \int_{0}^{\infty} \left( \frac{\delta S_1(f)}{S(f)} \right) \left( \frac{\delta S_2(f)}{S(f)} \right) df.
\label{eq:fisher_metric}
\end{equation}
The intrinsic instability of the detector, which we define as the geometric drift $\mathcal{D}(t)$, is the Fisher information velocity along the spectral trajectory $\dot{S}(t)$ \cite{KenningtonBlack2026_GTSP1}:
\begin{equation}
\mathcal{D}(t) = \frac{1}{2} \sqrt{g_S(\dot{S}, \dot{S})}.
\label{eq:fisher_velocity}
\end{equation}
A rigorous derivation of the underlying gauge-theoretic foundations and the principal bundle formalism governing this manifold is provided in Ref.~\cite{KenningtonBlack2026_GTSP1}.

\section{Methods}
\label{sec:methods}

\subsection{Dimensional Reduction and Basis Projection}
\label{subsec:dimensional_reduction}

To render the infinite-dimensional statistical manifold \cite{pistone1995infinite, lindquist2015geometric} computationally tractable for real-time analysis, we project the continuous Fisher information metric into a discrete frequency basis.
We establish a three-dimensional basis for operational deployment.
This specific projection maps the geometric vector directly onto the dominant physical noise mechanisms of the interferometer.
The low-frequency band, spanning 10 to 50 Hz, captures seismic up-conversion and control system noise below the interferometer's seismic wall \cite{matichard2015seismic, aasi2015advancedLIGO}.
The mid-frequency band, spanning 50 to 500 Hz, encompasses the primary astrophysical detection bandwidth where thermal and suspension noise dominate \cite{saulson1990thermal, martynovSensitivityAdvancedLIGO2016}.
The high-frequency band, spanning 500 to 2000 Hz, monitors variations in the quantum shot noise floor \cite{caves1981quantum, tseQuantumEnhancedAdvancedLIGO2019}.
These boundaries were chosen to align with the dominant physical noise mechanisms of the Advanced LIGO interferometers, though the framework supports arbitrary frequency partitioning.
By restricting the local logarithmic derivative of the PSD to these specific frequency intervals, we generate a discrete kinematic vector:
\begin{equation}
\vec{V}(t) = [V_{\text{low}}, V_{\text{mid}}, V_{\text{high}}].
\label{eq:kinematic_vector}
\end{equation}
The global scalar drift of the detector state, $\mathcal{D}(t)$, is subsequently computed as the $L_2$ norm of this kinematic vector, $||\vec{V}(t)||$ \cite{KenningtonBlack2026_GTSP1}.

\subsection{Vector Kinematics and Angular Evolution}
\label{subsec:vector_kinematics}

While the scalar drift norm measures the overall speed of the detector state evolution, the associated vector structure explicitly encodes the kinematics of the state change across the chosen basis.
To distinguish between proportional amplitude scaling and fundamental structural reconfigurations, we evaluate the tangent divergence of the tracking vector \cite{doCarmo1992riemannian, lee1997riemannian}.
We define the spectral warp indicator (Eq.~\ref{eq:spectral_warp_indicator}) by calculating the angular divergence between successive tangent vectors.
To ensure this calculation remains dimensionally agnostic and mathematically valid for any arbitrary $N$-dimensional frequency partitioning, we utilize exterior algebra \cite{spivak1999comprehensive, nakahara_geometry_2005}.
The sine of the divergence angle $\theta$ is computed via the magnitude of the wedge product between successive states, normalized by their respective scalar lengths:
\begin{equation}
\sin \theta \equiv \frac{||\vec{V}_t \wedge \vec{V}_{t+dt}||}{||\vec{V}_t|| \cdot ||\vec{V}_{t+dt}||}.
\label{eq:spectral_warp_indicator}
\end{equation}
In the deployed three-dimensional basis, this reduces to the familiar cross product magnitude $||\vec{V}_t \times \vec{V}_{t+dt}||$, though the exterior algebra formulation generalizes naturally to arbitrary $N$-dimensional frequency partitions.
This continuous angular measurement establishes a rigorous physical logic for anomaly classification.
A \textit{surge} event, characterized by $\sin \theta \approx 0$, represents a proportional scaling of the manifold where the spectral shape remains constant despite a broad amplitude fluctuation.
Conversely, a \textit{pivot} event, characterized by $\sin \theta \gg 0$, indicates a spectral warp, or differential redistribution of power across frequency bands, driven by a shifting physical noise mechanism.

\subsection{Software Implementation}
\label{subsec:software_implementation}

For operational deployment, we implemented this framework natively within the \texttt{sgn} stream-processing ecosystem \cite{huangSGNL2025}.
Standard block-based spectral estimation methods, such as Welch estimation \cite{welchUseFastFourier1967} or static median-geometric-mean algorithms \cite{littenberg_bayesline_2015}, require accumulating overlapping time-domain buffers, limiting the temporal resolution at which spectral evolution can be resolved.
To track manifold kinematics at the cadence required to resolve individual transients, our implementation utilizes a recursive streaming estimator \cite{haykin2002adaptive, KenningtonBlack2026_GTSP1, KenningtonBlack2026_GTSP2} rather than the median-geometric-mean algorithm used by the \texttt{sgnl} search pipeline \cite{huangSGNL2025}, which requires accumulating longer data buffers.
The primary state estimation utilizes a recursive exponential moving average with a smoothing coefficient $\alpha = 0.1$, applied to overlapping 1.0-second Fourier transforms with 50\% overlap, yielding an output cadence of 0.5 seconds.
At this cadence, the effective memory timescale of the estimator is approximately 5 seconds, providing sensitivity to both individual transients and slower instrumental drifts.
The \texttt{sgn-drift} package also provides a streaming median-geometric-mean implementation \cite{littenberg_bayesline_2015}. This more robust estimator is not used in the analyses presented here, but served to verify the accuracy of the recursive approach.
The spectral estimator feeds directly into the kinematic metric calculators.
The complete pipeline is published as the \texttt{sgn-drift} package\footnote{Source: \url{https://git.ligo.org/james.kennington/sgn-drift}}, which continuously ingests strain data, executes the band-limited integrations, computes the exterior algebra for the spectral warp indicator, and outputs the geometric state vector in real time.

\subsection{Data Selection}
\label{subsec:data_selection}

To validate the geometric framework across representative detector conditions, we selected eight segments of LIGO Livingston (L1) public strain data \cite{collaborationOpenDataLIGO2025, TheLIGOScientificCollaboration2015Advanced, capote_advanced_ligo_o4_2025} from the O4a observing run, totaling 39.3 hours of science-quality strain data.
The segments were chosen to span three qualitatively distinct noise regimes: three quiet periods with a stable noise floor, two transition periods exhibiting intermittent glitch activity \cite{abbottCharacterizationTransientNoise2016, davisLIGODetectorCharacterization2021, soni_ligo_detchar_o4a_2025}, and three scattering-active periods with elevated non-stationarity \cite{soniReducingScatteredLight2021}.
At the 0.5-second output cadence of the \texttt{sgn-drift} pipeline, these segments yield $N = 282{,}080$ independent manifold velocity samples spanning the full range of operational conditions encountered in practice.
This deliberate stratification ensures that the method is not evaluated on a single homogeneous dataset, but rather across the diversity of noise environments characteristic of the Advanced LIGO observing runs \cite{buikemaSensitivityPerformanceAdvanced2020, martynovSensitivityAdvancedLIGO2016}.
To validate veto safety, we additionally evaluated the geometric channel on 10 confirmed GWTC-4.0 astrophysical events with available L1 data \cite{collaborationGWTC40UpdatingGravitationalWave2025} and performed a systematic injection campaign of $4{,}852$ simulated compact binary coalescence signals.
Injected waveforms were generated using \texttt{IMRPhenomD} \cite{khan2016frequency} via \texttt{LALSimulation} \cite{lalsuite}, with component masses drawn randomly from $1$--$100\,M_\odot$ and optimal signal-to-noise ratios drawn log-uniformly from $5$ to $1{,}000$.
Injections were placed at randomized times within real O4a strain data spanning all three noise regimes.
We emphasize that the present analysis serves as a methodological demonstration on a single detector; a comprehensive multi-detector study spanning the full O4a dataset, including systematic comparison with established transient classification tools \cite{robinetOmicronTool2020, essickiDQStatistical2021, smithHierarchicalVeto2011}, is in preparation.

\section{Results}
\label{sec:results}

\subsection{Sensitivity Comparison with Standard Energy Monitors}
\label{subsec:sensitivity_comparison}

To quantify the sensitivity of the geometric channel relative to standard energy monitors, we compare the significance of anomalies detected by the Fisher velocity against those detected by BLRMS \cite{davis_improving_2019, davisLIGODetectorCharacterization2021}.
For each channel, we define the significance of an event as the MAD-normalized z-score $z = (x - \tilde{x}) / \sigma_{\text{MAD}}$, where $\tilde{x}$ is the per-segment median and $\sigma_{\text{MAD}}$ is the median absolute deviation.
This normalization ensures a fair comparison by calibrating each channel against its own noise floor.
Among the $N = 2{,}127$ top one-percentile manifold velocity events (with 10-second isolation) for which we computed paired BLRMS significance from coincident strain data, we evaluate two complementary results.

\begin{figure*}[htbp]
\centering
\includegraphics[width=0.9\textwidth]{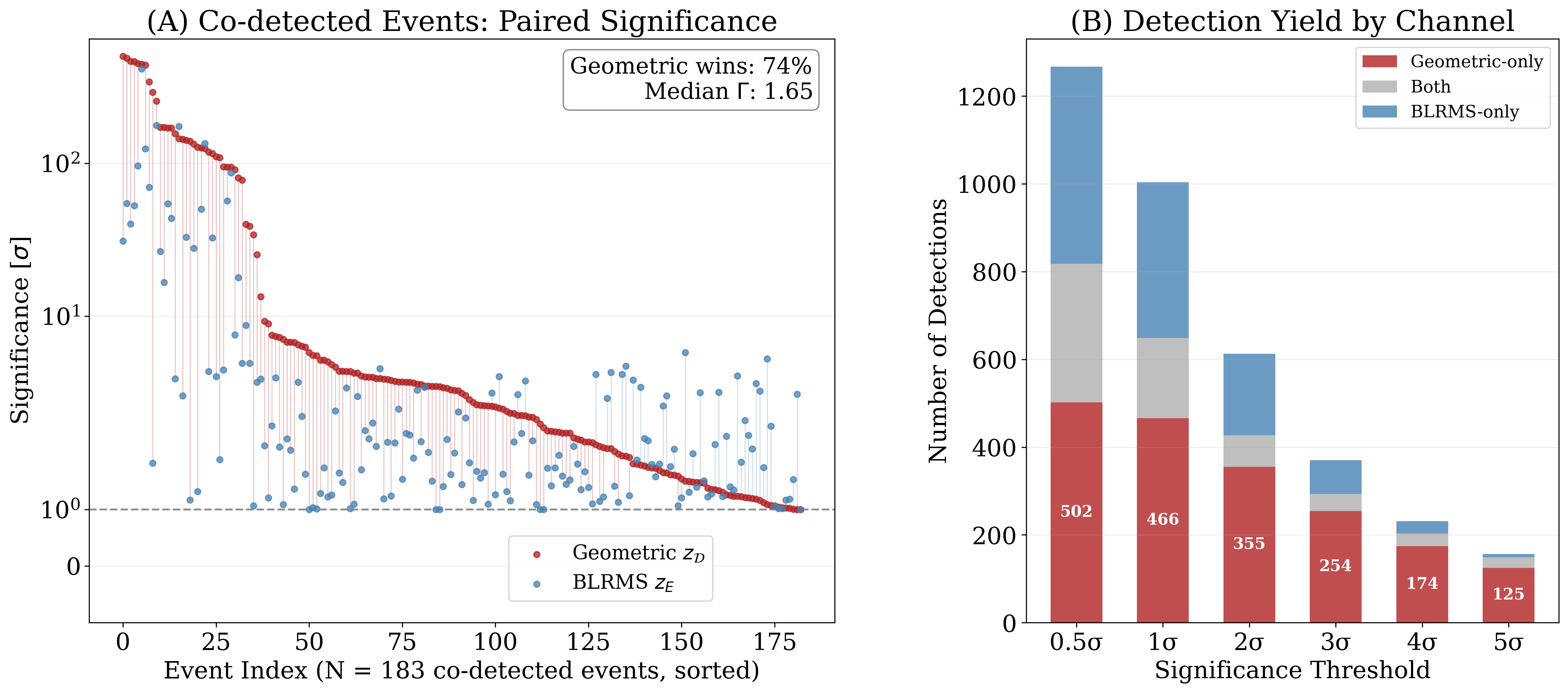}
\caption{Sensitivity comparison between the geometric channel and BLRMS. (A) Paired MAD-normalized significance ($z$-score) for $N = 183$ co-detected events at $1\sigma$, sorted by geometric significance. The geometric channel achieves higher significance in $74\%$ of cases with a median sensitivity ratio of $\Gamma = 1.65$. (B) Detection yield at matched significance thresholds, showing geometric-only (red), co-detected (grey), and BLRMS-only (blue) events. The geometric channel consistently identifies a larger unique population across all thresholds.}
\label{fig:sensitivity_comparison}
\end{figure*}

First, for events co-detected by both channels at $1\sigma$ significance ($N = 183$), we compute the sensitivity ratio
\begin{equation}
\Gamma \equiv \frac{z_{\mathcal{D}}}{z_{E}},
\label{eq:sensitivity_ratio}
\end{equation}
where $z_{\mathcal{D}}$ and $z_E$ are the geometric and energy significances, respectively.
The geometric channel achieves higher significance in $74\%$ of co-detected events (95\% CI via bootstrap resampling \cite{efronBootstrapMethods1979}: $[68\%, 80\%]$), with a median sensitivity ratio of $\Gamma = 1.65$ (95\% CI: $[1.49, 1.77]$).
This advantage strengthens at higher significance thresholds, reaching $92\%$ with median $\Gamma = 4.5$ at $3\sigma$ (Figure~\ref{fig:sensitivity_comparison}A).

Second, the two channels detect largely non-overlapping populations of anomalies.
At $1\sigma$ significance, the geometric channel identifies 466 unique events not flagged by BLRMS, while BLRMS identifies 355 unique events not flagged by the geometric channel, with only 183 events co-detected.
Adding the geometric channel to a BLRMS-only monitor increases the total anomaly catalog by $87\%$ (95\% CI: $[75\%, 96\%]$).
This complementarity increases with significance: at $3\sigma$, the geometric channel contributes 254 unique detections compared to 77 for BLRMS, with only 39 co-detected (Figure~\ref{fig:sensitivity_comparison}B).
The overall detection rate of $54$ severe events per hour is consistent with the $\mathcal{O}(10$--$100)$ triggers per hour typically produced by established transient classification tools \cite{robinetOmicronTool2020, essickiDQStatistical2021}.

\subsection{Statistical Taxonomy of Non-Stationarity}
\label{subsec:statistical_taxonomy}

\begin{figure}[htbp]
\centering
\includegraphics[width=0.9\columnwidth]{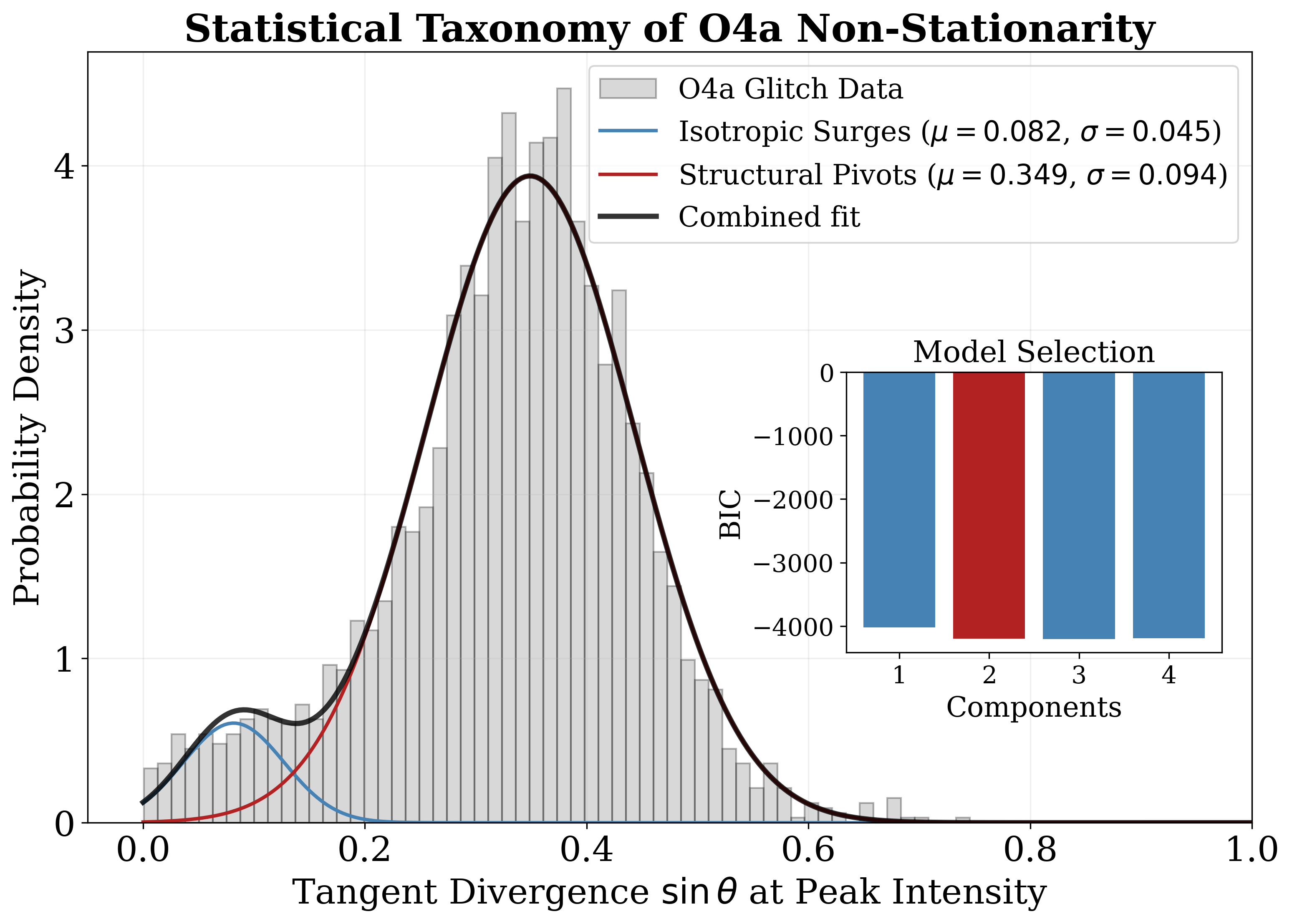}
\caption{Statistical taxonomy of O4a non-stationarity, revealing a bimodal distribution of tangent divergence ($\sin\theta$) at peak intensity. The low-divergence population (blue) corresponds to isotropic surges, while the dominant high-divergence population (red) corresponds to structural pivots.}
\label{fig:bimodal_taxonomy}
\end{figure}

To classify the physical nature of instrumental non-stationarity, we identified severe events by selecting all samples in the top one percent of the instantaneous manifold speed $||\vec{V}(t)||$ \cite{abbottCharacterizationTransientNoise2016}, yielding $2{,}686$ samples.
Gaussian mixture model selection via the Bayesian information criterion \cite{schwarzEstimatingDimensionModel1978} provides very strong evidence \cite{kassRafteryBayesFactors1995} for at least two distinct kinematic populations ($\Delta\text{BIC} = 178.5$ favoring $k=2$ over $k=1$), with only marginal evidence for additional sub-structure ($\Delta\text{BIC} = 5.9$ favoring $k=3$ over $k=2$).
The data resolves into two principal kinematic sub-populations (Figure~\ref{fig:bimodal_taxonomy}).
Among events detected by the geometric channel, the dominant population consists of structural pivots, representing $87.2\%$ (95\% CI: $[85.9\%, 88.4\%]$) with a mean tangent divergence of $\mu=0.362 \pm 0.003$.
These pivots correspond to spectral warps, in which the distribution of power across frequency bands is differentially restructured \cite{davisLIGODetectorCharacterization2021, cornish_bayeswave_2015}.
The secondary population consists of isotropic surges, making up the remaining $12.8\%$ (95\% CI: $[11.6\%, 14.1\%]$) of the events with a mean divergence of $\mu=0.119 \pm 0.006$.
These surges represent simple proportional intensity fluctuations that scale the spectral state without significantly altering its underlying structure.

\begin{figure*}[htbp]
\centering
\includegraphics[width=0.8\textwidth]{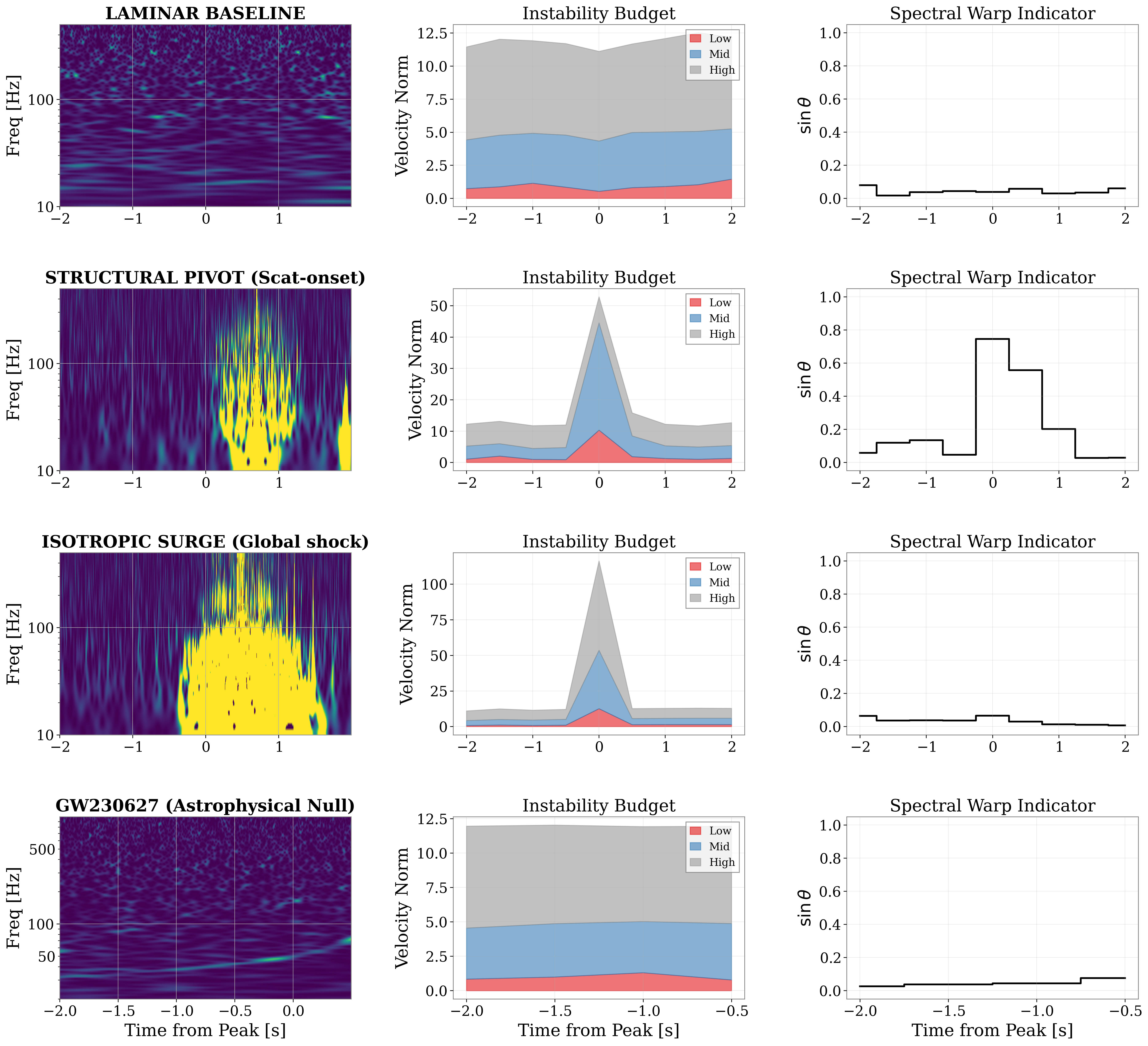}
\caption{Kinematic examples of the bimodal taxonomy. Each row shows (left) the spectrogram, (center) the instability budget decomposed by frequency band, and (right) the spectral warp indicator ($\sin\theta$). From top to bottom: a laminar baseline with low velocity and no warp, a structural pivot (scattering onset) with band-selective velocity increase and elevated $\sin\theta$, an isotropic surge (global shock) with high velocity but near-zero $\sin\theta$, and a confirmed astrophysical event (GW230627) showing no geometric channel response.}
\label{fig:taxonomy_comparison}
\end{figure*}

\subsection{Kinematic Orthogonality and Phase Space Mapping}
\label{subsec:kinematic_independence}

\begin{figure}[htbp]
\centering
\includegraphics[width=0.9\columnwidth]{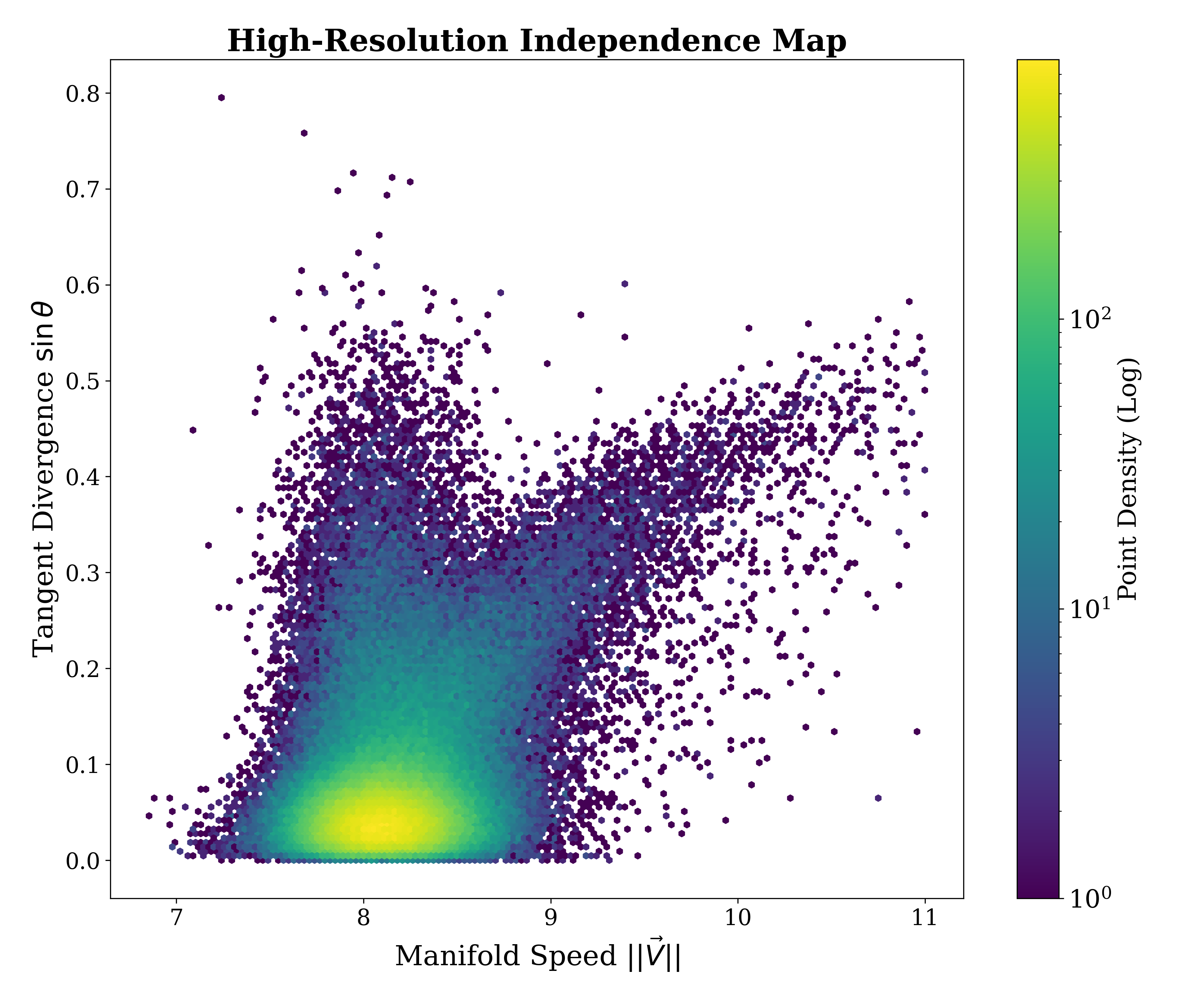}
\caption{High-resolution phase space map of tangent divergence ($\sin\theta$) versus manifold speed ($||\vec{V}||$), colored by point density on a logarithmic scale. The dense core at low speed represents background non-stationarity. At extreme velocities, the population bifurcates into distinct tail structures, confirming that speed and divergence encode independent physical information.}
\label{fig:phase_space_tails}
\end{figure}

To test the statistical independence of the kinematic variables, we map the tangent divergence against the absolute manifold speed in high resolution (Figure~\ref{fig:phase_space_tails}).
At low manifold speeds, the detector state wanders stochastically, forming a dense and unresolved core of background non-stationarity \cite{davisLIGODetectorCharacterization2021}.
However, at extreme velocities, the distribution does not follow a single continuum.
The population extends into distinct tail structures.
The vertical tail identifies events with high angular divergence at moderate speed, while the diagonal tail identifies events with both extreme velocity and significant spectral warp.
This separation confirms that manifold speed and tangent divergence encode independent physical information, validating their joint use as complementary diagnostics for automated data quality assessment \cite{abbottGuideLIGODetector2020}.
In practice, this independence means that a single scalar threshold on manifold speed alone would conflate surges and pivots, whereas the two-dimensional phase space enables targeted classification of the kinematic nature of each event.

\subsection{Veto Safety: The Astrophysical Null Result}
\label{subsec:veto_safety}

A viable detector characterization metric must remain insensitive to true astrophysical strain to avoid self-vetoing valid signals \cite{abbottGuideLIGODetector2020, veitch_robust_2015}.
We validate this property through two complementary analyses.

First, we evaluated the manifold drift during the transit of 10 confirmed O4a astrophysical events from the GWTC-4.0 catalog \cite{collaborationGWTC40UpdatingGravitationalWave2025} for which L1 public strain data was available \cite{collaborationOpenDataLIGO2025}.
In all 10 cases, the geometric channel registered a null result, with peak manifold velocities indistinguishable from the background median (maximum $z_{\mathcal{D}} = 1.26$, consistent with background fluctuations).
The response during GW230627 is shown in the final panel of Figure~\ref{fig:taxonomy_comparison}.

Second, we performed a systematic injection campaign, generating $N = 4{,}852$ simulated compact binary coalescence signals spanning the full parameter space of binary neutron star, neutron star--black hole, and binary black hole systems (component masses $1$--$100\,M_\odot$, optimal signal-to-noise ratios $5$--$1{,}000$) and injecting them into real O4a noise across quiet, transition, and scattering regimes.
In all $4{,}852$ injections, the Fisher velocity registered a null response, with a mean deviation from the no-injection baseline of $|\Delta| = 10^{-4}$, well below the detection threshold of $0.5$ used for anomaly identification. Even the single largest deviation ($|\Delta| = 0.21$) remained below this threshold.

This null response is expected: although a transient gravitational-wave signal contributes power to the strain, its contribution to the estimated PSD is negligible relative to the instrumental noise floor \cite{maggiore2007gravitational, creighton2011gravitational}, and the Fisher velocity tracks only fractional changes in $S(f)$.
The geometric channel therefore preferentially tracks instrumental state changes rather than astrophysical transients.

\section{Discussion}
\label{sec:discussion}

\subsection{Physical Interpretation of the Taxonomy}
\label{subsec:physical_interpretation}

The bimodal segregation of the noise manifold provides physical insight into the mechanisms of instrumental non-stationarity.
An isotropic surge represents a broad energy injection where the proportional distribution of noise across frequency bands remains stable.
Conversely, a structural pivot indicates a rapid alteration of the spectral composition, signifying that the physical mechanism has fundamentally warped the spectrum, driving the detector's trajectory on the noise manifold away from its prior course.
Our taxonomy analysis reveals that the vast majority ($87.2\%$) of severe events detected by the geometric channel involve spectral warping, while only approximately one in eight ($12.8\%$) are simple isotropic surges.
This distinction cannot be made by traditional scalar metrics, which collapse both phenomena into a single degenerate magnitude \cite{davisLIGODetectorCharacterization2021, abbottGuideLIGODetector2020}.

The two transients shown in the middle rows of Figure~\ref{fig:taxonomy_comparison} illustrate this degeneracy concretely.
Despite producing qualitatively similar broadband, sub-second excursions in the Q-transform spectrogram, the structural pivot and the isotropic surge are kinematically orthogonal: the former exhibits a sharp elevation in tangent divergence ($\sin\theta \approx 0.7$), indicating a differential redistribution of power across frequency bands, while the latter maintains $\sin\theta \approx 0$, consistent with a near-uniform amplitude scaling of the prior spectral state.
Submitted to the GravitySpy classifier \cite{zevinGravitySpyIntegrating2017}, both events are returned with the identical label \textit{Extremely Loud} at 100\% confidence, confirming that their distinction is not accessible from the spectrogram morphology alone.
This comparison illustrates the additional discriminatory information provided by the Fisher-geometric channel. While the two events appear sufficiently similar in their spectrogram morphology to be assigned the same GravitySpy label, the geometric representation distinguishes them through their markedly different spectral evolution.

\subsection{Complementarity with Energy-Based Monitors}
\label{subsec:complementarity}

The sensitivity comparison reveals that the geometric channel and standard BLRMS monitors detect largely non-overlapping populations of anomalies, with a Jaccard overlap of only $0.18$ at $1\sigma$ significance.
This complementarity is a direct consequence of the different physical quantities each channel measures: BLRMS tracks absolute acoustic power \cite{davis_improving_2019}, while the geometric channel tracks fractional spectral evolution.
Events that produce large absolute power fluctuations without altering the spectral shape are preferentially flagged by BLRMS, while events that restructure the noise floor without a large energy signature are preferentially flagged by the geometric channel.
The practical implication is that deploying both channels in parallel increases the total anomaly catalog by $87\%$ relative to BLRMS alone, substantially improving the completeness of detector characterization.
This low overlap also suggests that many spectral warps detected by the geometric channel would go entirely unnoticed by energy-based monitors, representing a previously uncharacterized class of instrumental non-stationarity.

\subsection{Operationalizing the Data}
\label{subsec:operationalizing_data}

Standard amplitude thresholding blindly vetoes all loud data, which needlessly reduces the operational duty cycle of the observatory \cite{buikemaSensitivityPerformanceAdvanced2020, martynovSensitivityAdvancedLIGO2016}.
The geometric channel enables targeted veto strategies based on the specific kinematic nature of the transient anomaly.
Because surges preserve the spectral shape, they are in principle more amenable to post-hoc correction strategies, such as dynamic amplitude rescaling, that could preserve the underlying data segment for search pipelines; however, the practical efficacy of such corrections remains to be demonstrated.
Conversely, severe spectral warps identified as pivots fundamentally corrupt the template inner product \cite{cutler_gravitational_1994, finn1992detection}, necessitating strict data quality vetoes.
By selectively vetoing only true manifold pivots, search pipelines can safely recover operational duty cycle and maximize astrophysical yield \cite{sachdev_gstlal_2019, nitz_pycbc_2018}.

\subsection{Veto Safety and Operational Deployment}
\label{subsec:veto_safety_deployment}

The comprehensive veto safety validation, encompassing 10 confirmed GWTC-4.0 events and $4{,}852$ simulated compact binary injections spanning the full parameter space, confirms that the geometric channel is intrinsically insensitive to astrophysical transients.
Combined with the demonstrated sensitivity advantage over BLRMS for co-detected events ($\Gamma = 1.65$) and the complementary detection of anomalies invisible to energy monitors, the geometric drift vector and tangent divergence angle represent strong candidates for automated data quality vetoes \cite{abbottGuideLIGODetector2020, soni_ligo_detchar_o4a_2025}.
Because the channel responds to fractional spectral changes rather than absolute power, it is naturally insensitive to the class of signals that matched-filter pipelines are designed to detect, reducing the risk of inadvertent signal loss from overly aggressive vetoing.

Utilizing the recursive estimation architecture of the \texttt{sgn-drift} package, this geometric channel is fully prepared for deployment as an online, low-latency control room monitor \cite{magee_first_2021}.
We note that the Fisher metric arises as the rigorous infinitesimal limit of the Bhattacharyya coefficient used in macroscopic pipelines such as PiNCh \cite{yarbrough_pinch_2025} (see Appendix~\ref{app:bhattacharyya_bridge}).
A comprehensive follow-up analysis, currently in preparation, will extend this framework across the full O4a observing run and all active detectors, including systematic cross-correlation with established transient classification tools such as Omicron \cite{robinetOmicronTool2020} triggers, iDQ \cite{essickiDQStatistical2021} scores, and hveto \cite{smithHierarchicalVeto2011} rounds, as well as characterization of the false positive rate under operational conditions.
Augmenting existing morphological and energy-based projections with continuous kinematic tracking provides the complementary precision required to more completely characterize current and next-generation detector networks \cite{punturo2010einstein, reitze2019cosmic}.

\section{Conclusion}
\label{sec:conclusion}

This work establishes Fisher information velocity as a new geometric channel for precision glitch identification in gravitational-wave detectors \cite{abbottCharacterizationTransientNoise2016, davisLIGODetectorCharacterization2021}.
By modeling the PSD as a point on a Riemannian manifold \cite{lee1997riemannian, amariMethodsInformationGeometry2000}, this framework decouples simple acoustic scaling from physical reconfigurations of the instrumental noise floor, a distinction inaccessible to traditional scalar metrics.
The resulting geometric channel is not a replacement for existing energy monitors but a fundamentally complementary tool: it detects a largely independent population of anomalies and provides a physically motivated surge/pivot taxonomy that enables targeted veto strategies.
Its demonstrated insensitivity to astrophysical transients makes it a safe candidate for automated deployment alongside existing data quality infrastructure \cite{abbottGuideLIGODetector2020}.
Future work will focus on scaling the \texttt{sgn-drift} streaming pipeline to process the complete O4a observing run across all active detectors, generating a comprehensive anomaly catalog with systematic cross-correlation against established transient classification tools \cite{robinetOmicronTool2020, essickiDQStatistical2021, smithHierarchicalVeto2011}.
Continuous geometric tracking of the spectral manifold offers a new operational dimension for stabilizing detection pipelines and maximizing astrophysical yield in current and next-generation observatory networks \cite{punturo2010einstein, reitze2019cosmic}.

\begin{acknowledgments}
This material is based upon work supported by NSF's LIGO Laboratory which is a major facility fully funded by the National Science Foundation.
This research has made use of data, software and/or web tools obtained from the Gravitational Wave Open Science Center (https://www.gw-openscience.org/), a service of LIGO Laboratory, the LIGO Scientific Collaboration, and the Virgo Collaboration.
The authors acknowledge support from the National Science Foundation under awards OAC-2103662, PHY-2308881, PHY-2011865, OAC-2201445, OAC-2018299, PHY-0757058, PHY-0823459, PHY-2207728, PHY-2513124, PHY-2110594, and PHY-2513358.
The authors are grateful for computational resources provided by the Pennsylvania State University's Institute for Computational and Data Sciences gravitational-wave cluster, and the LIGO Lab cluster at the LIGO Laboratory.
\end{acknowledgments}

\appendix

\section{Connection to Macroscopic Pipelines}
\label{app:bhattacharyya_bridge}

The continuous geometric formulation presented in this work is directly connected to established macroscopic anomaly detection pipelines, such as PiNCh (Pipeline-Informed Noise Characterization) \cite{yarbrough_pinch_2025}.
Such pipelines evaluate non-stationarity over time by utilizing the Bhattacharyya coefficient \cite{bhattacharyyaMeasureDivergence1943} to calculate the overlap between discrete probability distributions of search triggers.
These discrete distributions are typically separated by macroscopic observational chunks, denoted by $\Delta t$.

\begin{figure}[htbp]
\centering
\includegraphics[width=\columnwidth]{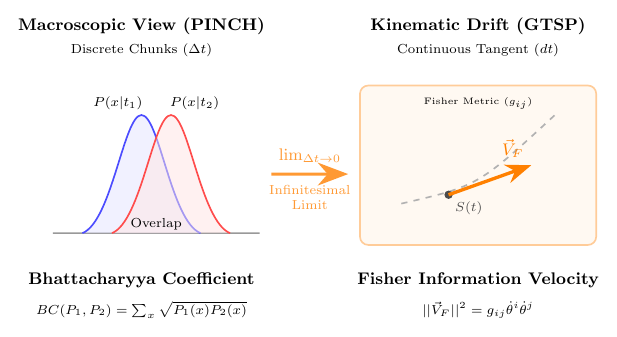}
\caption{The mathematical bridge demonstrating the continuous Fisher information metric as the infinitesimal limit of the macroscopic Bhattacharyya coefficient.}
\label{fig:bhattacharyya_bridge}
\end{figure}

If we shrink the observation window from macroscopic chunks down to a continuous infinitesimal stream as $\Delta t \to 0$, the statistical overlap converges exactly into a geometric derivative \cite{amariMethodsInformationGeometry2000, cover2006elements}.
Expanding the Bhattacharyya coefficient to second order in the fractional perturbation $\delta S / S$ recovers the Fisher metric (Eq.~\ref{eq:fisher_metric}) as the leading-order term \cite{cover2006elements, amariMethodsInformationGeometry2000}, establishing it as the rigorous infinitesimal limit of the Bhattacharyya distance.
Therefore, while macroscopic tools measure the discrete secant of spectral divergence, our geometric framework tracks the instantaneous tangent of that exact same physical evolution.

\bibliography{references}

\end{document}